\RequirePackage{fix-cm}
\documentclass[twocolumn,epjc3]{svjour3}      
\RequirePackage{graphicx}
\RequirePackage{mathptmx}      
\RequirePackage{flushend}
\RequirePackage[numbers,sort&compress]{natbib}
\RequirePackage[colorlinks,citecolor=blue,urlcolor=blue,linkcolor=blue]{hyperref}
\RequirePackage[T1]{fontenc}
\journalname{Eur. Phys. J. C}
\smartqed  
\newcommand\beq{\begin{eqnarray}}
\newcommand\eeq{\end{eqnarray}}
\newcommand\bq{\begin{equation}}
\newcommand\eq{\end{equation}}
\begin{document}
\title{Probing neutrino nature at Borexino detector with chromium neutrino source}
\author{W. Sobk\'ow\thanksref{e1,addr1} \and A. B\l{}aut\thanksref{e2,addr1} 
}
\thankstext{e1}{e-mail: wieslaw.sobkow@ift.uni.wroc.pl}
\thankstext{e2}{e-mail: arkadiusz.blaut@ift.uni.wroc.pl}

\institute{ Institute of Theoretical Physics, University of Wroc\l{}aw,
Pl. M. Born 9, PL-50-204~Wroc{\l}aw, Poland\label{addr1}
}

\date{Received: date / Accepted: date}

\maketitle
\begin{abstract}

In this paper, we indicate a possibility of utilizing the intense chromium source ($\sim 370\; PBq$) in probing the neutrino nature in low energy neutrino experiments with the ultra-low threshold and background real-time Borexino detector located near the source ($\sim 8\;m$). We analyze the elastic scattering of  electron neutrinos (Dirac or Majorana, respectively) on the unpolarized electrons in the relativistic neutrino limit. We assume that the incoming neutrino beam is the  superposition of left-right chiral states. Left chiral neutrinos may be detected by the standard $V - A$ and non-standard  scalar $S_L$, tensor $T_L$ interactions, while right chiral ones partake only  in the exotic $V + A$ and $S_R, T_R$ interactions. 
Our model-independent study is carried out for the flavour (current) neutrino eigenstates.   We compute the expected event number for the standard $V-A$ interaction of the left chiral neutrinos using the current experimental values of standard couplings and  in the case of left-right chiral superposition. We show that the significant decrement in the event number due to the interference terms between the standard  and exotic interactions for the Majorana $\nu_e$'s   may appear.  The $90 \% \; C. L. $ sensitivity contours in the planes of corresponding  exotic couplings  are also found. The presence of interferences in the Majorana case give the stronger constraints than for the Dirac neutrinos, even if the neutrino source is placed outside the detector.
\end{abstract}
\section{Introduction} 
\label{sec1}
Possibility of utilizing  various artificial neutrino sources (ANS) in the low energy neutrino  ($\nu$) experiments with the ultra-low background and threshold detectors to explore the Lorentz structure of weak interactions and other non-standard $\nu$ properties has been discussed in many papers, e. g.  
\cite{Vogel,Ferrari,Ianni,Ianni1,Miranda}. As is well known there are essentially  two types of the ANS which can be used in the large liquid scintillator detectors; the monochromatic $\nu_e$ emitters   $(e. g. ^{51}Cr, ^{37}Ar, ^{49}V, ^{145}Sm)$, and $\overline{\nu}_e $ sources with continuous $\beta$ spectrum   (e. g. $^{144}Cs - ^{144}Pr, ^{106}Ru - ^{106}Rh, ^{90}Sr-^{90}Y$, $^{42}Ar - ^{42}K $) \cite{Korn,sterile7}.  
$^{51}Cr$ source as the dichromatic neutrino emitter with energies of $430 \;keV $ $ (10 \;\%)$  and $750\;keV$ $(90 \;\%)$, and a 
mean life time $(\tau \simeq  40 $  days)  has already been  utilized to calibrate GALLEX and SAGE experiments \cite{Gallex,Gallex1,Gallex2,Sage,Sage1,Sage2}, where a  deficit in the rate of $\nu$  interactions has been found \cite{Bahcall,Giunti}. Presently, the $^{51}Cr$ emitter with   activity of the order of $\sim 370 \; PBq \;(\sim 10 \; MCi)$  in the SOX experiment (Short distance Oscillation with boreXino) with the Borexino detector will be used to search for the sterile $\nu_e$'s \cite{sterile7,sterile,sterile1,sterile2,sterile3,sterile4,sterile5,sterile6} and to improve the current limits on the neutrino magnetic moment 
\cite{Beda}, and to reduce the uncertainty on the direct measurement of the standard couplings. 
 It is worthy of reminding that the extremely low background  Borexino detector has precisely measured the low energy solar $\nu_e$ components ($^{7}Be, pep$) \cite{Be,Be1,pep} and detected the geophysical $\overline{\nu}_{e}$'s \cite{geo}. \\
This detector seems to be an appropriate tool to test the $\nu$ nature, i. e. whether  $\nu$'s are  the Dirac or Majorana fermions. 
The problem of distinguishing between the Dirac and Majorana  $\nu$'s   can be investigated  in the context of non-vanishing $\nu$ mass and  of standard vector-axial $(V-A)$ weak interaction of the only left chiral (LCh) $\nu$'s, using purely leptonic processes such as the polarized muon decay at rest  or the mentioned   neutrino-electron elastic scattering (NEES). Kayser \cite{Kayser} and Langacker  
\cite{Langacker} have proposed the first tests concerning the mass dependence, however it is worthwhile noting the other papers devoted to the various aspects of  $\nu$ nature, e. g.  \cite{Zralek,Nishiura,Semikoz,Pastor,Barranco,Singh,Gutierrez}. 
It is necessary to point out that the current experiments regarding the discrimination between the  Dirac and Majorana  $\nu$'s  are mainly based on the searching for  the neutrinoless double beta decay (NDBD) \cite{Majorana}, however  the low energy $\nu$ experiments with the intense ANS, very low background and threshold detector seem to have  similar scientific opportunities, and  may also shed some light on this problem. 
It is important to emphasize that there is also an alternative scenario  within the relativistic $\nu$ limit, when  one departs from the $V-A$ interaction and  one admits the  exotic scalar $(S)$, tensor $(T)$, pseudoscalar $(P)$ and $(V+A)$ weak interactions  of the right chiral (RCh) $\nu$'s (right-handed helicity when $m_{\nu} \rightarrow 0$) in the leptonic  processes. 
 The proper tests have been reported by Rosen \cite{Rosen} and Dass \cite{Dass}. 
 It is relevant to  remark that  the existing data  still leaves a little space for the exotic couplings of the interacting RCh $\nu$'s outside the SM \cite{SM}. 
Let us recall that the SM  does not clarify  the origin of  parity violation (PV)  at current energies. It is well known that  the SM PV is incorporated in ad hoc way by assuming that gauge boson couples only to the left chiral currents.  However on the other hand,  there is no experimental evidence of the parity conservation at higher energies so far. Moreover,  the SM does not explain the observed baryon asymmetry of universe \cite{barion} through  a single CP-violating phase of the Cabibbo-Kobayashi-Maskawa quark-mixing matrix (CKM) \cite{Kobayashi}, the large hierarchy fermion mass\-es, and  other  fundamental  aspects.  Consequently, a lot of  non-standard schemes with the Majorana (and Dirac) $\nu$'s, time reversal violation (TRV) exotic interactions, mechanisms explaining the origin of fermion generations, masses, mixing and smallness of $\nu$ mass appeared.  It is worthwhile to mention the  non-standard $\nu$ interactions  (NSI) changing and conserving $\nu$ flavour 
\cite{NSI}, which may be generated by the mechanisms of massive neutrino models \cite{massivenu}. The NSI  phen\-omen\-ology has been extensively explored  
\cite{fenNSI}. Concerning interacting RCh  $\nu$'s, the suitable non-standard models seem to be  the left-right symmetric models (LRSM) \cite{Pati},  composite models (CM, where tensor and scalar interactions are generated by the exchange of constituents) \cite{Jodidio, CM}, models with extra dimensions (MED) 
\cite{Extra}, the unparticle  models (UP) \cite{unparticle}. In the MED the LCh standard particles  live on the three-brane, while the RCh  $\nu$'s can move in the extra dimensions. It causes that  the interactions of RCh  $\nu$'s  with the LCh fermions are extremely tiny to be observed. It is worthy of stressing that in the UP scheme  the leptons with  the different chiralities   can couple to the spin-0 scalar, spin-1 vector, spin-2 tensor unparticle sectors. It means that the amplitude for 
NEES  can have the form of the unparticle four-fermion contact interaction at low energies, and contain the exotic  contributions.   
Currently there is no unambiguous indication of new non-standard  gauge model, because the experimental possibilities are still limited. There is a constant  necessity of improvement of the  precision of  present tests at low energies, and on the other hand, the precise measurements of new observables including the linear  terms from  the exotic couplings would be required. 
\\ In this study, we concentrate on  the application of  $^{51}Cr$ electron neutrino $(\nu_e)$ source deployed at $8.25\;m$ from the  centre of Borexino detector to find the allowed limits on the exotic $S$, $T$, $V+A$ couplings in the relativistic $\nu_e$ limit, when the incoming $\nu_e$ beam  is the  superposition of left-right chiral states and has Dirac or Majorana nature.  
We analyze  the elastic  scattering of  $\nu_e$ beam  off the unpolarized electron target as the detection process of possible exotic signals. It should be pointed out that the scintillator detector does not allow to observe the directionality of the recoil electrons, so all the interference terms between the standard and exotic couplings in the differential cross section vanish for the Dirac $\nu_e$'s, and only the contributions from the squares of exotic couplings of the RCh  $\nu$'s and of non-standard couplings of LCh ones (and at most  the interferences within exotic couplings) may generate the possible effect. The situation is distinct for the Majorana $\nu$'s, where some linear terms coming from the exotic couplings after the integration over the azimuthal angle of outgoing electron momentum may occur.    
One of the goals is to show in model-independent way how the expected event number  for the standard $V-A$ interaction depends on the precision of measurement of standard  couplings. Next, we calculate  the predicted event number coming from  the admixture of exotic interactions both for the Dirac and Majorana  
$\nu_e$'s.  Finally, we find the $90 \,\% \, C. L. $ sensitivity contours in the planes of proper  exotic couplings for both scenarios. 
\section{Elastic scattering of Dirac electron neutrinos  off unpolarized electrons }
\label{sec2}
We  assume that the incoming monochromatic  Dirac $\nu_{e}$  beam comes from the electron capture by $^{51}Cr$  $(e^{-} + ^{51}Cr \rightarrow \nu_e + 
^{51}V)$ and  is  the superposition of left-right chiral states. LCh $\nu_{e}$'s 
 are  mainly detected by the standard $V-A$ interaction, while RCh ones are detected only by the exotic scalar $S$, tensor $T$, $V+A$ interactions in the elastic scattering on the unpolarized electrons; $\nu_{e} + e^{-}\rightarrow \nu_{e} +e^{-}$. The considered scenario admits also the detection of $\nu_{e}$'s  with left-handed chirality by the non-standard $S$ and $T$ interactions. It is important to  emphasize that   our analysis is carried out for the flavour (current) $\nu_e$ eigenstates. The amplitude for the   $\nu_{e}  e^{-}$ scattering   at low energies takes the form:
\beq \label{ampD} M^{D}_{\nu_{e} e^{-}}
&=&
\frac{G_{F}}{\sqrt{2}}\{(\overline{u}_{e'}\gamma^{\alpha}(c_{V}^{L}
- c_{A}^{L}\gamma_{5})u_{e}) (\overline{u}_{\nu_{e'}}
\gamma_{\alpha}(1 - \gamma_{5})u_{\nu_{e}})\nonumber\\ 
&& \mbox{} + (\overline{u}_{e'}\gamma^{\alpha}(c_{V}^{R}
+ c_{A}^{R}\gamma_{5})u_{e}) (\overline{u}_{\nu_{e'}}
\gamma_{\alpha}(1 + \gamma_{5})u_{\nu_{e}})  \\
&  & \mbox{} +
c_{S}^{R}(\overline{u}_{e'}u_{e})(\overline{u}_{\nu_{e'}}
(1 - \gamma_{5})u_{\nu_{e}}) \nonumber\\
&& \mbox{} +
\frac{1}{2}c_{T}^{R}(\overline{u}_{e'}\sigma^{\alpha \beta}u_{e})(\overline{u}_{\nu_{e'}}
\sigma_{\alpha \beta}(1 - \gamma_{5})u_{\nu_{e}})\nonumber\\
&&\mbox{} +
c_{S}^{L}(\overline{u}_{e'}u_{e})(\overline{u}_{\nu_{e'}}
(1 + \gamma_{5})u_{\nu_{e}}) \nonumber\\
&& \mbox{} +
\frac{1}{2}c_{T}^{L}(\overline{u}_{e'}\sigma^{\alpha \beta}u_{e})(\overline{u}_{\nu_{e'}}
\sigma_{\alpha \beta}(1 + \gamma_{5})u_{\nu_{e}})
\}\nonumber,  
 \eeq
 where $G_{F} = 1.1663788(7)\times
10^{-5}\,\mbox{GeV}^{-2} (0.6 \; ppm)$ \cite{Mulan} is the Fermi constant. 
  The coupling constants are
denoted with the superscripts $L $ and $R $ as $c_{V}^{L, R} $, 
$c_{A}^{L, R}$, $c_{S}^{R, L}$, $c_{T}^{R, L}$ respectively to the incoming $\nu_{e}$ 
of left- and right-handed chirality.  Because we take into account  the TRV, all the coupling constants  are complex. 
It is worthy of pointing out that we probe the case when the outgoing electron direction
is not observed, so the laboratory differential
cross section is presented after integration over the azimuthal
angle $\phi_{e}$ of the recoil electron momentum. The obtained formula, in the relativistic limit, does not contain the interference terms between  the standard $c_{V, A}^{L}$ and exotic $c_{S,T}^{L, R}, c_{V, A}^{R} $ couplings:   
\bq \label{przekrnue} 
 \frac{d\sigma}{d y_{e} } = \bigg(\frac{d \sigma}{d y_{e}}\bigg)_{(V- A)} + \bigg(\frac{d \sigma}{d y_{e}}\bigg)_{(V+A)}  
+ \bigg(\frac{d \sigma}{d y_{e} }\bigg)_{(S, T)}, \nonumber  
\eq 
\beq \label{VA} 
\lefteqn{\bigg( \frac{d \sigma}{d
y_{e} } \bigg)_{(V- A)} = B \bigg\{ (1-\mbox{\boldmath
$\hat{\eta}_{\nu}$}\cdot\hat{\bf q}
) \bigg[|c_{V}^{L} + c_{A}^{L}|^{2}}\\
&& \mbox{} + |c_{V}^{L} - c_{A}^{L}|^{2}(1-y_{e})^{2}
 - \frac{m_{e}y_{e}}{E_{\nu}}\left(|c_{V}^{L}|^{2} - |c_{A}^{L}|^{2}\right) \bigg] \bigg\},
\nonumber \eeq 
\beq \lefteqn{\bigg(\frac{d \sigma}{d y_{e} }\bigg)_{(V+A)} = B \bigg\{ (1+\mbox{\boldmath
$\hat{\eta}_{\nu}$}\cdot\hat{\bf q}
) \bigg[|c_{V}^{R} + c_{A}^{R}|^{2} }\\
&& \mbox{} + |c_{V}^{R} - c_{A}^{R}|^{2}(1-y_{e})^{2}
 - \frac{m_{e}y_{e}}{E_{\nu}}\left(|c_{V}^{R}|^{2} - |c_{A}^{R}|^{2}\right) \bigg] \bigg\},  
\nonumber \eeq
\beq \lefteqn{\bigg(\frac{d \sigma}{d y_{e}}\bigg)_{(S, T)} = \mbox{}
B\Bigg\{ (1+\mbox{\boldmath $\hat{\eta}_{\nu}$}\cdot\hat{\bf
q})\bigg[ \frac{1}{2}y_{e}\left(y_{e}+2\frac{m_{e}}{E_{\nu}}\right)
 |c_{S}^{R}|^{2} } \nonumber \\
&& \mbox{} + \left((2-y_{e})^2 -\frac{m_{e}}{E_{\nu}}y_{e}\right)|{c_{T}^{R}}|^{2}
+ y_{e}(y_{e}-2)Re(c_{S}^{R}c_{T}^{*R}) \bigg] \nonumber\\
&& \mbox{} +  (1-\mbox{\boldmath $\hat{\eta}_{\nu}$}\cdot\hat{\bf
q})\bigg[ \frac{1}{2}y_{e}\left(y_{e}+2\frac{m_{e}}{E_{\nu}}\right) |c_{S}^{L}|^{2} \\
&& \mbox{} + \left((2-y_{e})^2 -\frac{m_{e}}{E_{\nu}}y_{e}\right)|{c_{T}^{L}}|^{2}
+ y_{e}(y_{e}-2)Re(c_{S}^{L}c_{T}^{*L}) \bigg] \Bigg\}, \nonumber\\
y_{e} & \equiv &
\frac{T_e}{E_\nu}=\frac{m_{e}}{E_{\nu}}\frac{2 cos^{2}\theta_{e}}
{(1+\frac{m_{e}}{E_{\nu}})^{2}-cos^{2}\theta_{e}} \eeq 
is the ratio of the
kinetic energy of the recoil electron $T_{e}$  to the incoming $\nu_e$ 
energy $E_{\nu}$; 
$\theta_{e}$ is the angle between  the direction of the outgoing electron
momentum  $ \hat{\bf p}_{e}$ and  $\nu_{e}$ 
LAB momentum unit vector ${\bf \hat{q}}$  (recoil electron scattering angle); $m_{e}$ is the
electron mass; $B\equiv \left( E_{\nu} m_{e}/2\pi\right)\left(G_{F}^{2}/2\right)$;  $\mbox{\boldmath $\hat{\eta}_{\nu}$}$ is the unit 3-vector of
 $\nu_{e}$ spin  polarization in its rest frame; $(\mbox{\boldmath$\hat{\eta}_{\nu}$}\cdot\hat{\bf q}){\bf\hat{q}}$ is the longitudinal component of $\nu_{e}$ spin polarization; 
$|\mbox{\boldmath $\hat{\eta}_{\nu}$}\cdot\hat{\bf q}| = |1- 2
Q_{L}^{\nu}|$; $Q_{L}^{\nu}$ is the probability of producing the LCh $\nu_{e}$.  \\ 
 It can be noticed that there are only the contributions from 
the $T$-even longitudinal component of the  $\nu_{e}$ spin polarization, and   no linear terms from the exotic couplings  in the relativistic limit appear. 
  The  formula on the number of events  for the standard  and non-standard interactions is similar as in \cite{Ianni}: 
\beq 
\label{ne}
\lefteqn{N = N_t \cdot \Phi_0 \cdot \Gamma(t_{tr}, t_{ex}) \cdot F(h=\frac{R}{D})\cdot \int_{ 250 \; keV}^{ 700 \; keV} d T_{e}^{m}\cdot} \\
&& \mbox{}  \bigg[ 0. 81\int_{0}^{T_{e}^{max}(E_\nu=746 \; keV)} d T_e R(T_{e}^{m}, T_e)\frac{d \sigma}{d T_{e}}\mid_{E_\nu=746 \; keV} \nonumber \\
&& \mbox{} + 0. 09 \int_{0}^{T_{e}^{max}(E_\nu=751 \; keV)} d T_e  R(T_{e}^{m}, T_e)\frac{d \sigma}{d T_{e}}\mid_{E_\nu=751 \; keV}\bigg]\nonumber.
\eeq
In order to compute the expected  event number for the standard $V-A$ interaction, we use the experimental values of standard couplings: $c_{V}^{L}= 1 + (-0.04 \pm 0.015), c_{A}^{L}= 1+ \\(-0.507 \pm 0.014)$   \cite{Data}.
Assumptions concerning the technical setup are analogical as in \cite{Ianni} except the stronger source activity.  $N_e= 3.3 \cdot 10^{32}$  is  the number of electrons calculated for $100$ tons of  spherical fiducial volume $(R= 3\;m)$ of the detector; $D=8. 25 \; m$ is the distance between the chromium source and detector centre; $\Phi_0=(I_0/4\pi D^{2})$; $I_0= 370 \; PBq= 10 MCi$ is the intensity of the source at end of bombardment; 
$F(h=R/D)= (3/2h^3)
\{h - [(1-h^2)/2]\; ln[(1+h)/(1-h)]\}$\\ $\simeq 1.028$ is the factor taking into account the geometry of the system; $\Gamma(t_{tr}, t_{ex})= \tau exp(-t_{tr}/\tau )[1-exp(-t_{ex}\tau )]$;\\ $\tau=(T_{1/2}/ln2)\simeq 39. 97$ days;  $t_{tr}=5$ days; $t_{ex}=60$ days define the exposure time. 
\beq  
R(T_{e}^{m}, T_e)&=& \frac{1}{\sqrt{2\pi} \delta(T_e)} exp\left[-\frac{(T_{e}^{m}-T_e)^2}{2\delta^{2}(T_e)}\right]
\eeq
 is the detector resolution function; $\delta(T_e)/keV = 48 \sqrt{T_{e}/MeV}$ is the  electron energy resolution; $T_{e}^{m}\in[250, 700]\; keV$ is the energy window for the reconstructed recoil electron kinetic energy.\\ 
Fig.1 shows how the uncertainty on the measurement of   standard $c_{V,A}^{L}$ couplings  affects the expected event number. In this case we assume pure LCh $\nu_e$ beam with $\mbox{\boldmath$\hat{\eta}_{\nu}$}\cdot\hat{\bf q}=-1$. 
Fig. 2 illustrates the dependence of event number on $\mbox{\boldmath$\hat{\eta}_{\nu}$}\cdot\hat{\bf q} \in [-1, 1]$ for the standard interaction (solid thick line) and various combinations of exotic couplings (other lines).  
Fig. 3 demonstrates the predicted event number coming from the superposition of left-right chiral $\nu_e$'s for two chosen  scenarios $(c_{V}^{L}$, $c_{A}^{L}$, $c_{V}^{R}$, $c_{A}^{R})$, $(c_{V}^{L}$, $c_{A}^{L}$, $c_{S}^{R}$, $c_{T}^{R})$ (upper plots) and one for the pure LCh $\nu_e$ beam (lower plot). We use the experimental values for the standard couplings $c_{V}^{L}=1 -0.04$, $c_{A}^{L}=1-0.507$ and 
probe the interval $[-0.6,\;0.6]$ of all the exotic couplings. 
It is important to stress that for the left-right chiral superposition of $\nu_e$ states $\mbox{\boldmath$\hat{\eta}_{\nu}$}\cdot\hat{\bf q}\not=-1$. In order to illustrate all possible effects from the exotic interactions of RCh $\nu_e$'s we assume $\boldmath\hat{\eta}_{\nu}\cdot\hat{\bf q}=-0.75$ corresponding to $P_L=0.875$. For the scenario with only LCh $\nu_e$'s participating both in the standard 
$V-A$ and non-standard $S_L,T_L$ interactions we take $\boldmath\hat{\eta}_{\nu}\cdot\hat{\bf q}=-1$. 
Fig. 4 illustrates $90\%\;C.L.$ sensitivity contours in the planes ($c_{V}^{R}$, $c_{A}^{R}$), ($c_{S}^{R}$, $c_{T}^{R}$), ($c_{S}^{L}$, $c_{T}^{L}$), respectively. It is worth noting that we consider six degrees of freedom and then carry out the projection onto the appropriate plane of couplings. The proper contours are calculated with the use of inequality taken from \cite{Ianni}: 
\beq
\label{CLD}
\left| \frac{N}{N_{SM}} -1\right|&\geq& \epsilon_{90}=3.263 \frac{\delta N_{SM}}{N_{SM}}, 
\eeq
where $ \delta N_{SM}= \sqrt{N_{B} + N_{SM}(1 + N_{SM} \delta_{A}^{2})}$ is the total $1 \sigma$ uncertainty of the signal; $\delta_{A}=0.01$ is the uncertainty of  source activity. We assume that the number of background events is $N_{B}=4380$ as in \cite{Ianni}.  
\begin{figure}
\begin{center}
\includegraphics[scale=.6]{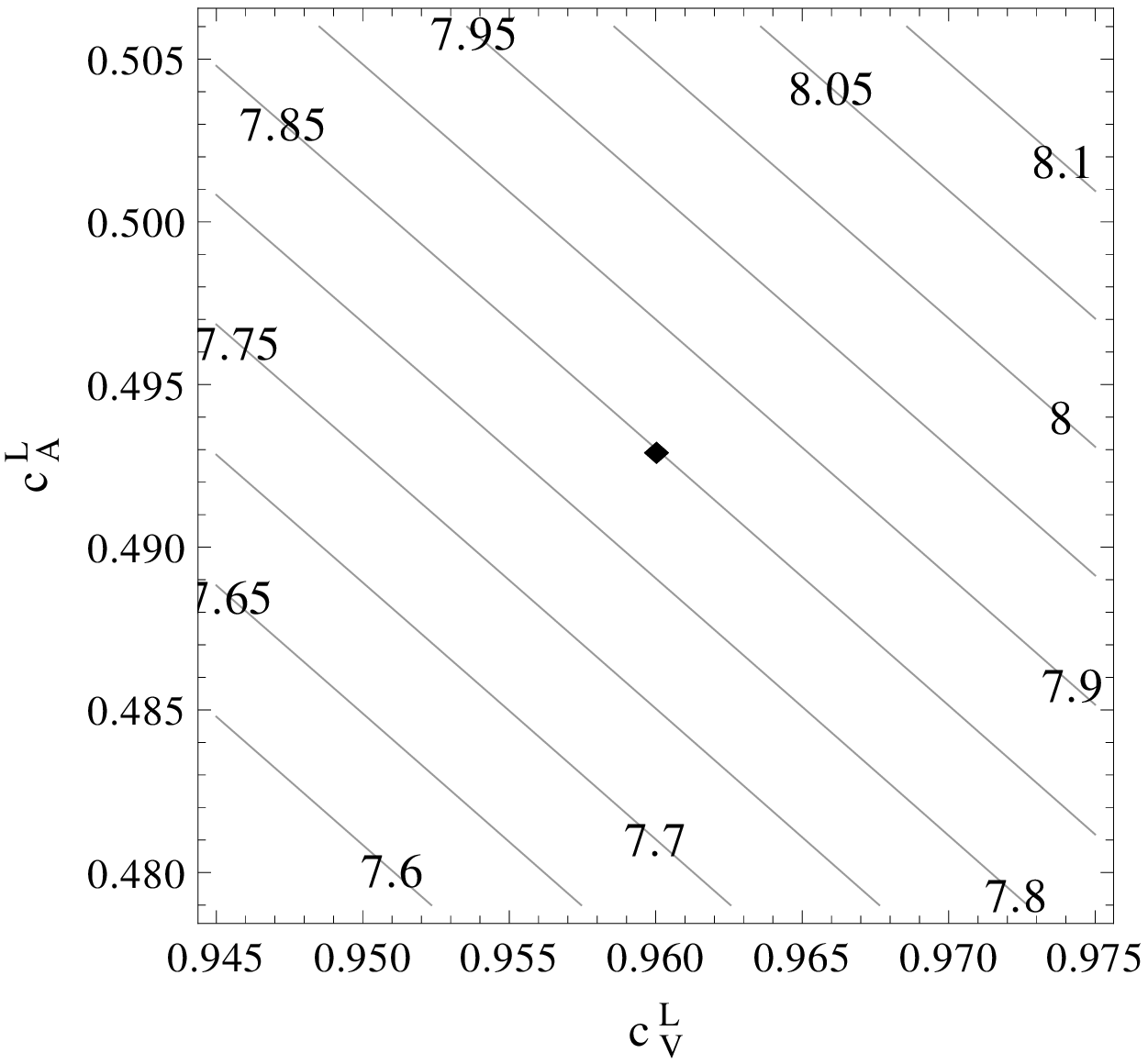}
\end{center}
\caption{Dirac $\nu_e$. Dependence of the event number $N/10^3$ on the errors of the experimental 
values of the standard $c_{V}^{L}, c_{A}^{L}$ couplings.  \label{Fig1D}}
\end{figure}

\begin{figure}
\begin{center}
\includegraphics[scale=.6]{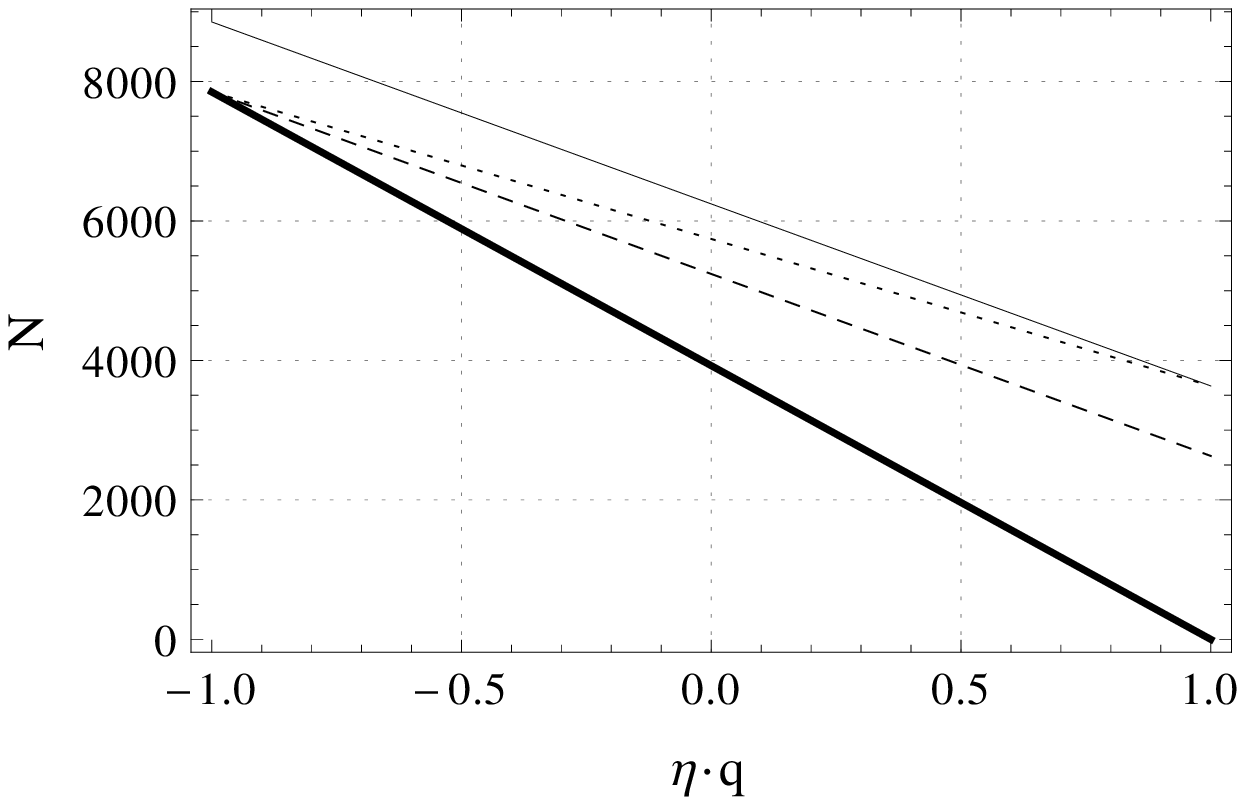}
\end{center}
\caption{Dirac $\nu_e$. Dependence of the event number on 
$(\boldmath\hat{\eta}_{\nu}\cdot\hat{\bf q}) \in [-1, 1]$. 
Solid thick line for the pure standard $V-A$ interaction;
dashed line, dotted line and thin line for combinations of the standard $V-A$ interaction 
with
($c_{V}^{R}=c_{A}^{R}=0.4,\;c_{S}^{R}=c_{T}^{R}=c_{S}^{L}=c_{T}^{L}=0$),
($c_{V}^{R}=c_{A}^{R}=c_{S}^{R}=c_{T}^{R}=0.4,\;c_{S}^{L}=c_{T}^{L}=0$)
and 
($c_{V}^{R}=c_{A}^{R}=c_{S}^{R}=c_{T}^{R}=c_{S}^{L}=c_{T}^{L}=0.4$)
respectively. \label{Fig2D}}
\end{figure}

\begin{figure}
\begin{center}
\includegraphics[scale=.6]{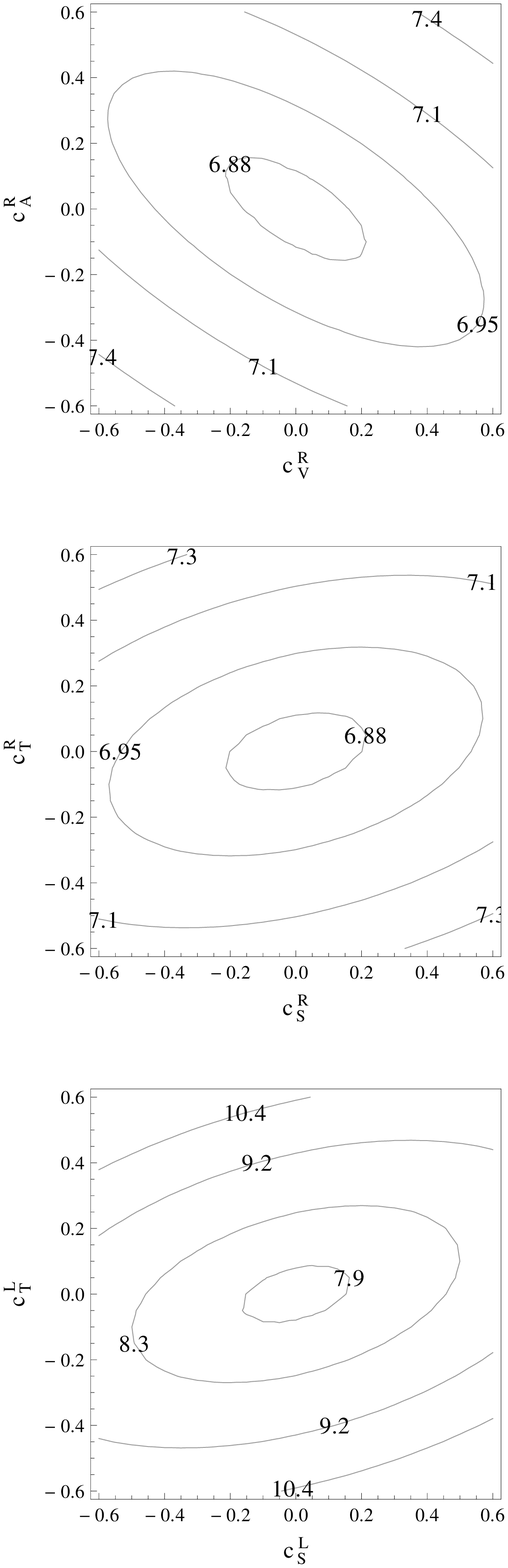}
\end{center}
\caption{Dirac $\nu_e$. Two upper plots show the predicted event number $N/10^3$ coming from the superposition of left-right chiral $\nu_e$'s for two   scenarios $(c_{V}^{L}, c_{A}^{L}, c_{V}^{R}, c_{A}^{R})$, $(c_{V}^{L}, c_{A}^{L}, c_{S}^{R}, c_{T}^{R})$, respectively with $\mbox{\boldmath$\hat{\eta}_{\nu}$}\cdot\hat{\bf q}=-0.75$. Lower plot concerns  $(c_{V}^{L}, c_{A}^{L}, c_{S}^{L}, c_{T}^{L})$ interactions of  LCh $\nu_e$'s   with $\mbox{\boldmath$\hat{\eta}_{\nu}$}\cdot\hat{\bf q}=-1$. \label{Fig3D}}
\end{figure}
\begin{figure}
\begin{center}
\includegraphics[scale=.7]{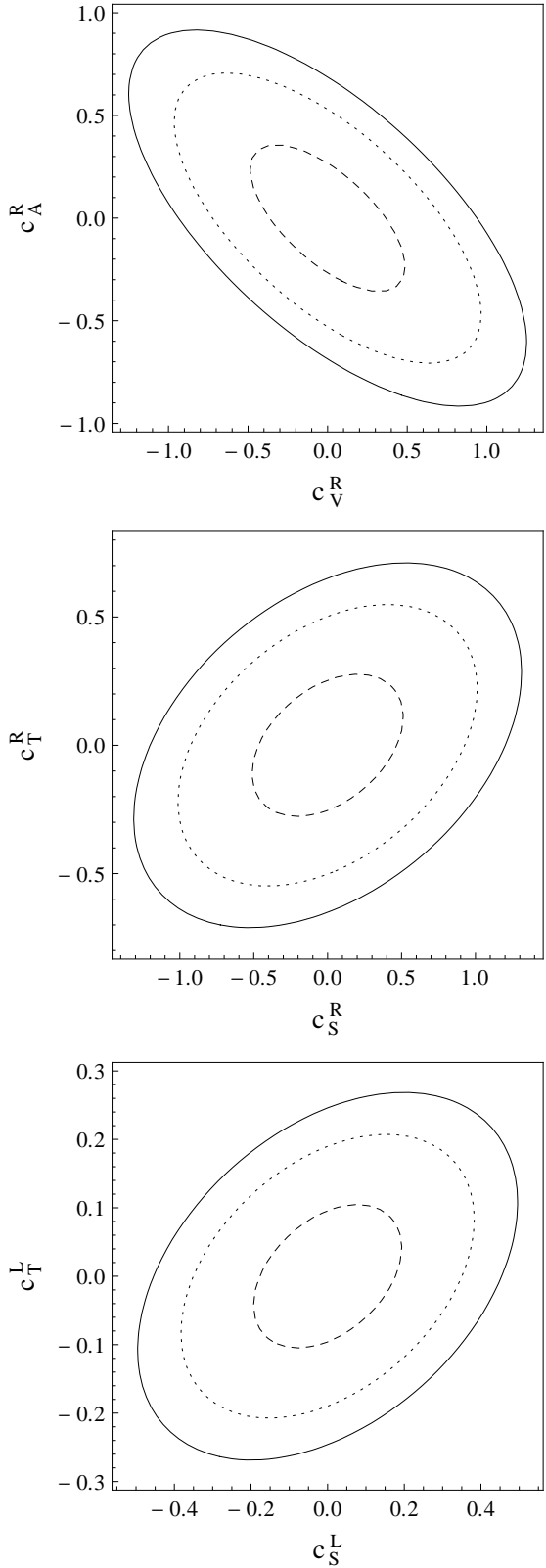}
\end{center}
\caption{Dirac $\nu_e$. $90\%\;C.L.$ sensitivity contours in the planes ($c_{V}^{R}$, $c_{A}^{R}$), ($c_{S}^{R}$, $c_{T}^{R}$) for $\boldmath\hat{\eta}_{\nu}\cdot\hat{\bf q}=-0.75$,  
and the plane ($c_{S}^{L}$, $c_{T}^{L}$) with $\boldmath\hat{\eta}_{\nu}\cdot\hat{\bf q}=-1$,  
respectively.  Dashed line of each plot is for the source located at the detector centre and for $\delta_{A}=0.001$. Dotted line is for $\delta_{A}=0.01$ and the chromium source at the detector centre. Solid line is for $D=8.25\;m$ and $\delta_{A}=0.01$.  \label{Fig4D}}
\end{figure}


\section{Elastic scattering of Majorana electron  neutrinos off unpolarized electrons} 
\label{sec3}
The amplitude for the elastic  scattering of  the Majorana  $\nu_{e}$'s on the unpolarized electrons  at low energies has the form (one assumes the flavour  $\nu_e$ eigenstates similarly as for the Dirac case): 
\beq \label{ampM} \lefteqn{ M^{M}_{\nu_{e} e^{-}}
  =  \mbox{} \frac{2G_{F}}{\sqrt{2}}\{-(\overline{u}_{e'}\gamma^{\alpha}(c_{V}
- c_{A}\gamma_{5})u_{e}) (\overline{u}_{\nu_{e'}}\gamma_{\alpha}\gamma_{5} u_{\nu_{e}}) }\\ 
 &+& (\overline{u}_{e'}\gamma^{\alpha}(\tilde{c}_{V}
+ \tilde{c}_{A}\gamma_{5})u_{e}) (\overline{u}_{\nu_{e'}}\gamma_{\alpha}\gamma_{5} u_{\nu_{e}})  \nonumber \\
& + &  c_{S}^{L}(\overline{u}_{e'}u_{e})(\overline{u}_{\nu_{e'}}
(1 - \gamma_{5})u_{\nu_{e}}) +  c_{S}^{R}(\overline{u}_{e'}u_{e})(\overline{u}_{\nu_{e'}}(1 + \gamma_{5})u_{\nu_{e}})\}.  \nonumber 
 \eeq
 One can  see that the neutrino part  of the above amplitude does not contain the contribution from the  $V$  and $T$  interactions in contrast to the Dirac case, where both terms partake.  The  $V+A$ interaction  is also admitted.  Moreover, the  $A$ and $S$  contributions are multiplied by factor 2 and  this is a direct consequence of the fact that the Majorana neutrino is described by the self-conjugate field. The indexes $L$, ($R$) for $c_V, c_A \; (\tilde{c}_V, \tilde{c}_A)$  couplings are omitted. It means that both LCh and RCh $\nu_{e}$'s may participate in the standard A and non-standard $\tilde{A}$ interactions of Majorana $\nu_{e}$'s off the electron target. 
The exotic $S$ coupling constants are 
denoted with the superscripts $L $ and $R $ as  $c_{S}^{L, R}$ respectively to the incoming $\nu_{e}$ 
of left- and right-handed chirality.   All the couplings are assumed to be complex numbers as for the Dirac case. 
 The differential cross section for the elastic  scattering of Majorana current $ \nu_{e}$'s on the unpolarized electrons in the relativistic limit  has the form:  
 \beq \label{przekMaj} \frac{d \sigma}{d y_{e}} = \bigg( \frac{d \sigma}{d y_{e}}\bigg)_{(V- A)} &+& \bigg( \frac{d \sigma}{d y_{e}}\bigg)_{( \tilde{V}+ \tilde{A})} \\
\mbox{} + \bigg( \frac{d \sigma}{d y_{e} }\bigg)_{(V-A)( \tilde{V}+ \tilde{A})} 
 &+& \bigg( \frac{d \sigma}{d y_{e}}\bigg)_{(S_{L}, S_{R})}, \nonumber
\nonumber \eeq 
\beq \label{MAL} \lefteqn{\bigg( \frac{d \sigma}{d y_{e}} \bigg)_{(V- A)} = \mbox{}  B \bigg\{ 
- \frac{2 m_{e}y_{e}}{E_{\nu}}\left(|c_{V}|^{2} - |c_{A}|^{2}\right)} \\
&& \mbox{} + |c_{V} + c_{A}|^{2} (2+ (1+\mbox{\boldmath
$\hat{\eta}_{\nu}$}\cdot\hat{\bf q})(y_{e}-2)y_{e})\nonumber\\
&& \mbox{} + |c_{V} - c_{A}|^{2} (2 + (1-\mbox{\boldmath
$\hat{\eta}_{\nu}$}\cdot\hat{\bf q})(y_{e}-2)y_{e})
\bigg\},\nonumber\eeq 
\beq \label{MAR} \lefteqn{\bigg( \frac{d \sigma}{d y_{e}} \bigg)_{( \tilde{V}+ \tilde{A})} = 
\mbox{}  B \bigg\{ - \frac{2 m_{e}y_{e}}{E_{\nu}}\left(|\tilde{c}_{V}|^{2} - |\tilde{c}_{A}|^{2}\right)
 }\nonumber\\
&& \mbox{} + |\tilde{c}_{V} - \tilde{c}_{A}|^{2} (2+ (1+\mbox{\boldmath
$\hat{\eta}_{\nu}$}\cdot\hat{\bf q})(y_{e}-2)y_{e})\nonumber\\
&& \mbox{} + |\tilde{c}_{V} + \tilde{c}_{A}|^{2} (2+ (1-\mbox{\boldmath
$\hat{\eta}_{\nu}$}\cdot\hat{\bf q})(y_{e}-2)y_{e})\bigg\},\eeq
\beq
\label{MALMAR} \lefteqn{\bigg( \frac{d \sigma}{d y_{e}}\bigg)_{(V-A)(\tilde{V}+ \tilde{A})} }\nonumber\\
&=& \mbox{}
 4 B \bigg\{  Re(c_{A}\tilde{c}_{A}^{*})\bigg(2+ (y_{e}-2)y_{e} + \frac{ m_{e}y_{e}}{E_{\nu}}\bigg) \\
&& \mbox{} + \bigg(Re(c_{V}\tilde{c}_{A}^{*})- Re(c_{A}\tilde{c}_{V}^{*})\bigg) (y_{e}-2)y_{e} \mbox{\boldmath $\hat{\eta}_{\nu}$}\cdot\hat{\bf q} \nonumber\\ 
 && \mbox{} + Re(c_{V}\tilde{c}_{V}^{*})\bigg(-2 - (y_{e}-2)y_{e}  + \frac{ m_{e}y_{e}}{E_{\nu}}\bigg)  \bigg\}, \nonumber\eeq
 \beq
\label{MSLR} \lefteqn{\bigg(\frac{d  \sigma}{d y_{e}}\bigg)_{(S_L,S_R)} = \mbox{}
B\bigg\{  2 y_{e}\left(y_{e}+2\frac{m_{e}}{E_{\nu}}\right)}\\
&& \mbox{} \cdot \bigg[ (1+\mbox{\boldmath $\hat{\eta}_{\nu}$}\cdot\hat{\bf q})|c_{S}^{R}|^{2} 
 + (1-\mbox{\boldmath $\hat{\eta}_{\nu}$}\cdot\hat{\bf q})|c_{S}^{L}|^{2}\bigg]\bigg\}.\nonumber\eeq
 
We see that the interference terms between  ($c_V$, $c_A$) and  ($\tilde{c}_V$, $\tilde{c}_A$) couplings appear in contrast to the Dirac case, where such contributions annihilate. It can also be noticed that the differential cross section does not contain  $T$-odd observables similarly as for the Dirac $\nu_e$'s. 
It is obvious that the predicted  event number for the pure $V-A$ interaction in the Majorana case is the same as for the Dirac $\nu_e$'s.  
However if one departs from the standard couplings and allows for the exotic interactions, the possibility of distinguishing between the Dirac and Majorana  
$\nu$'s in the limit o vanishing $\nu_e$ mass due to the  interferences appears. Fig. 5 shows how the event number depends on 
$\boldmath\hat{\eta}_{\nu}\cdot\hat{\bf q} \in [-1, 1]$ in the standard case and for the exotic interactions. It is noteworthy that the presence of interference terms may cause a decrease of event number (thin line) in contrast to the Dirac scenario, where such a regularity is impossible. Fig. 6 illustrates the predicted event number for the superposition of left-right chiral $\nu_e$'s in the case of two scenarios. Upper plot shows clearly the impact of the interferences between  
($c_{V}$, $c_{A}$) and ($\tilde{c}_V$, $\tilde{c}_A$) on the event number for given $\boldmath\hat{\eta}_{\nu}\cdot\hat{\bf q}=-0.75$.  The significant decrement in the event number in comparison with the Dirac case It can be noticed. Fig. 7 demonstrates $90\%\;C.L.$ sensitivity contours in the planes  ($\tilde{c}_V$, $\tilde{c}_A$), ($c_{S}^{L}$, $c_{S}^{R}$), respectively. In the present case we admit forth degrees of freedom for the Majorana $\nu_e$'s (inequality (\ref{CLD}) with  $\epsilon_{90}=2.789 (\delta N_{SM}/N_{SM})$ and then carry out the projection onto the appropriate plane of couplings. 
One can  see the qualitative difference due to the interferences in comparison with the Dirac case, even when the source is placed at $D=8.25\;m$  with $\delta_{A}=0.01$ (solid line). 

\begin{figure}

\begin{center}
\includegraphics[scale=.5]{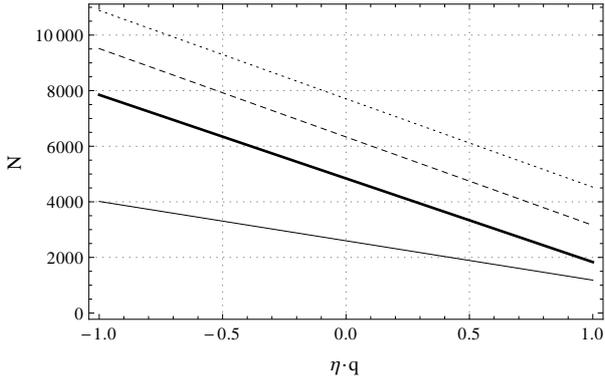}
\end{center}
\caption{Majorana $\nu_e$.Dependence of the event number on 
$(\boldmath\hat{\eta}_{\nu}\cdot\hat{\bf q}) \in [-1, 1]$. 
Solid thick line for the pure standard $V-A$ interaction;
dashed line, dotted line and thin line for combinations of the standard $V-A$ interaction 
with
($\tilde{c}_{V}=\tilde{c}_{A}=0.4,\;c_{S}^{L}=c_{S}^{R}=0$),
($\tilde{c}_{V}=\tilde{c}_{A}=c_{S}^{L}=c_{S}^{R}=0.4$)
and 
($\tilde{c}_{V}=0.2,\;\tilde{c}_{A}=-0.2,\;c_{S}^{L}=c_{S}^{R}=0.4$)
respectively.  \label{Fig5M}}
\end{figure}
\begin{figure}
\begin{center}
\includegraphics[scale=.6]{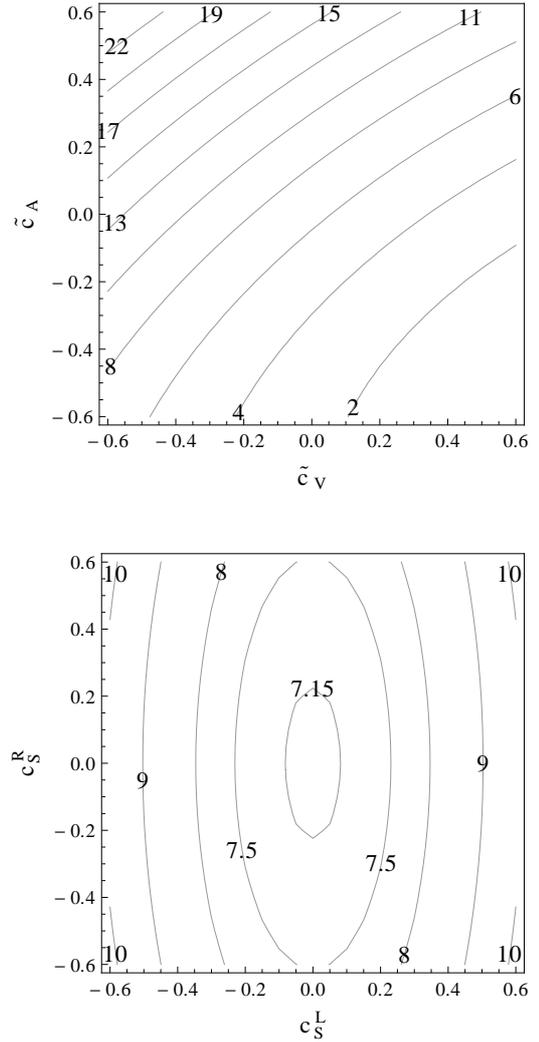}
\end{center}
\caption{Majorana $\nu_e$. Upper plot shows the expected event number $N/10^3$ for nonzero
($c_{V}$, $c_{A}$, $\tilde{c}_V$, $\tilde{c}_A$) couplings; 
the interferences between ($c_{V}$, $c_{A}$) and ($\tilde{c}_V$, $\tilde{c}_A$)
can significantly decrease the event number. Lower plot is for nonzero ($c_{V}$, $c_{A}$, $c_{S}^{L}$, $c_{S}^{R}$). Both cases with $\boldmath\hat{\eta}_{\nu}\cdot\hat{\bf q}=-0.75$. \label{Fig6M}}
\end{figure}
\begin{figure}
\begin{center}
\includegraphics[scale=.7]{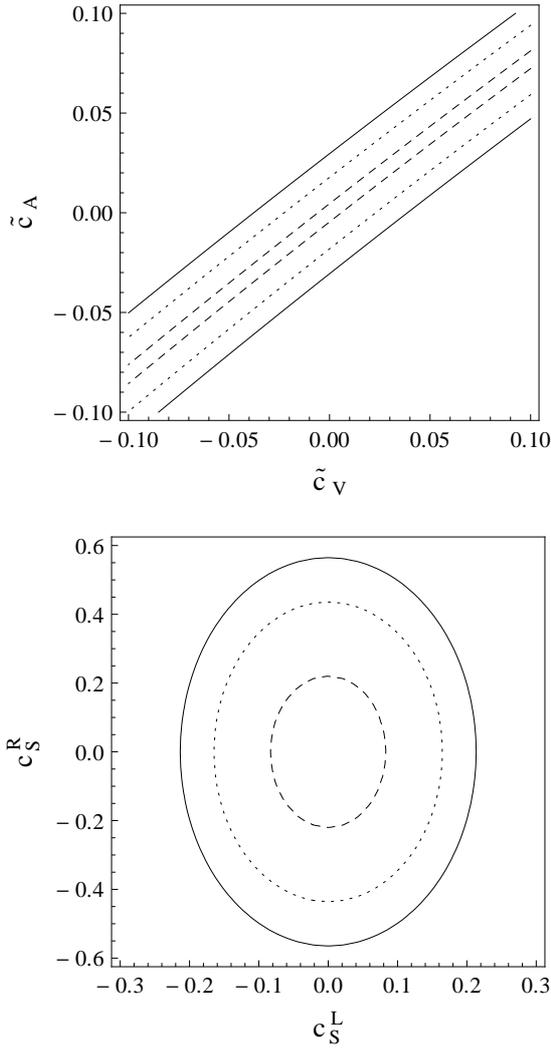}
\end{center}
\caption{Majorana $\nu_e$. $90\%\;C.L.$ sensitivity contours in the planes  ($\tilde{c}_V$, $\tilde{c}_A$), ($c_{S}^{L}$, $c_{S}^{R}$) for $\boldmath\hat{\eta}_{\nu}\cdot\hat{\bf q}=-0.75$.  
Dashed line of each plot is for the source located at the detector centre and $\delta_{A}=0.001$. Dotted line is for $\delta_{A}=0.01$ and the chromium source at the detector centre. 
Solid line for $D=8.25\;m$ and $\delta_{A}=0.01$. \label{Fig7M}}
\end{figure}



\section{Conclusions}
We have shown that the high-precision low-energy  experiment with the intense $^{51}Cr$ $\nu_e$ source located at near distance from the ultra-low threshold Borexino detector centre may be useful tool to test the $\nu$ nature problem in the limit of relativistic $\nu$. It is important to stress that the interference terms between $c_{V, A}$ and $\tilde{c}_{V,A}$  couplings for the Majorana $\nu_e$'s do not vanish. It means that even if the  intense $^{51}Cr$ source is deployed outside the Borexino detector, the significant decrement in the event number caused by the mentioned interferences   may occur. Such a regularity in the Dirac case does not manifest (no linear contributions from the exotic couplings survive).  Although the chromium source can not be placed at the detector centre, such a location would allow  more sensitive tests of the exotic couplings and of the $\nu$ nature, provided that the errors on the  activity source  are very tiny.  As is known the beta emitter $(^{144}Cs - ^{144}Pr)$  with the deployment at the detector centre is considered, so the combined analysis for both sources would  constrain  stringently the allowed region on the exotic couplings and  shed more light on the fundamental question of $\nu$ nature (in preparation).  
  
\label{sec4}

\end{document}